\documentclass[twoside]{photon2007}
\usepackage[latin1]{inputenc}
\usepackage[dvips]{graphicx,epsfig,color}
\usepackage{wrapfig,rotating}
\usepackage{amssymb,amsmath,array}

\begin{document}
\title{
Photoproduction in Ultra-Peripheral \\ Heavy-Ion Collisions}
\author{Joakim Nystrand
\vspace{.3cm}\\
Department of Physics and Technology, \\ 
University of Bergen, Bergen, Norway 
}

\maketitle

\begin{abstract}
The strong electromagnetic fields present in ultra-peripheral collisions 
of heavy-ions offer a possibility to study two-photon and photonuclear 
collisions complementary to similar studies with lepton beams but over
an increased photon energy range. 
This presentation\cite{jnystrand} will give an overview of photoproduction 
at hadron colliders. 
\end{abstract}

\section{Introduction}

The idea to use the strong electromagnetic fields in 
high energy proton-proton and nucleus-nucleus collisions to 
study photon-induced interactions has attracted an increased 
interest in recent years. Various aspects of such interactions have  
been discussed in presentations at earlier conferences in this 
series during the last ten 
years\cite{Photon05,Photon03,Photon01,Photon99,Photon97}, and it 
is gratifying to see that an entire session is devoted to the topic this year. 

This presentation will give an overview of photoproduction at 
hadron colliders. The focus will be on the accelerators with the 
highest collision energy in the world currently: 
the Relativistic Heavy Ion Collider (RHIC) at 
Brookhaven National Laboratory (proton-proton and heavy-ion collisions), 
The Fermilab Tevatron (proton-anti-proton collisions), and the 
CERN Large Hadron Collider (first proton-proton collisions expected in 2008, 
later also heavy-ion collisions). More complete reviews can be found in 
Refs.~\cite{Baltz:2007kq,Bertulani:2005ru}.

\section{Photoproduction at hadron and electron colliders} 

Photon-induced interactions have been studied with lepton beams in 
fixed target experiments and at colliders. But photon-hadron
and two-photon interactions can occur also when the lepton-beams are 
replaced by ultrarelativistic protons or heavy nuclei. The maximum photon 
energies are then restricted by the form factor of the projectile. 
An additional restriction on the ``useful'' photon energy spectrum comes 
from the requirement that there be no other, hadronic interactions in 
the same event. The presence of a form factor, when translated into impact
parameter space, roughly corresponds to a cut on the minimum impact parameter 
of $b > R$, where $R$ is the radius of the projectile, while the 
requirement that there be no accompanying hadronic interactions is more 
restrictive and corresponds to $b > 2R$. 
The presence of a form factor also means that only interactions with 
low-virtuality photons can be studied. 
Despite these limitations, the very high 
collision energies of current hadron colliders lead to 
useful photon energies exceeding those at HERA and LEP, for example. 

Since the intensity of the electromagnetic field is proportional to 
the square of the charge of the beam particle, photon-induced
interactions are enhanced by up to a factor $Z^2$ (Z is the atomic 
number of the ion) in heavy-ion collisions.

The spectrum of equivalent photons in pp and heavy-ion collisions are 
shown in Fig.~1. The luminosity scale on the y-axis should be interpreted 
with care, in particular for pp collisions, since at the maximum beam 
luminosities there are usually several overlapping events in a single 
bunch-crossing. Overlapping events may preclude a clean separation of the 
photon-induced reactions and the effective photon luminosities might 
therefore be lower. 

A major difference between lepton- and hadron-beams is that with 
hadron beams a certain reaction channel can often proceed both via 
a strong interaction as well as through an electromagnetic interaction.
One illustrative example is the production of two (heavy) quarks. 
At hadron colliders, this can happen through a purely 
electromagnetic (two-photon) process, 
$\gamma + \gamma \rightarrow Q \overline{Q}$. It can also proceed via 
photon-gluon fusion $\gamma + g \rightarrow Q \overline{Q}$ and 
gluon-gluon fusion $g + g \rightarrow Q \overline{Q}$. 
The two-gluon fusion production is the dominant production mechanism for 
pairs of $b$ and $c$ quarks. Estimates show 
that the ratios of the cross sections for the three processes in 
Pb+Pb interactions at the LHC are 
roughly $1$:$10^3$:$10^6$\cite{Nystrand:2006gi}.  
This is a consequence of 
the different coupling strengths (strong vs. electromagnetic) and the cut-offs 
introduced by the form factors. 
The cross section for purely electromagnetic production of a pair of 
heavy quarks is thus only a fraction $\sim 10^{-6}$ of the cross section for the 
dominating production mode through two-gluon fusion.
The large difference in  
cross section is one reason why special trigger and analysis techniques are needed 
to study photon-induced processes at hadron colliders. 

One should note, however, that some ultra-peripheral reaction channels 
have very high cross sections, in some cases even higher than the total 
hadronic cross section. This is the case for two-photon production of 
$e^+ e^-$-pairs and exclusive production of $\rho^0$-mesons in 
high energy heavy-ion collisions.

\begin{figure}[tbh]
\centerline{\includegraphics[width=1.0\columnwidth]{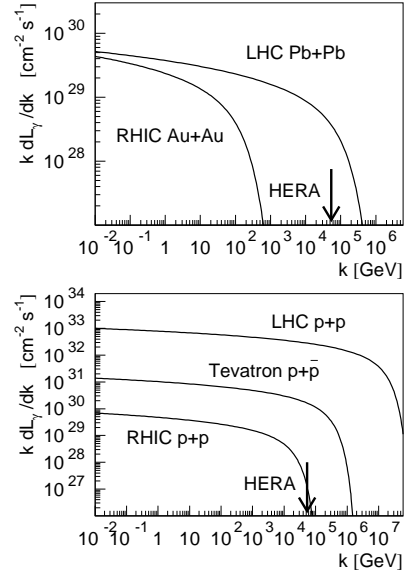}}
\caption{Equivalent photon luminosity (the photon spectrum $dn/dk$ 
multiplied by the beam luminosity) in heavy-ion (top) and proton-proton 
collisions (bottom). The photon energy, $k$,  is calculated in the rest frame 
of one of the projectiles. The arrow shows the maximum photon energy in 
$e+p$ collisions at HERA. The spectra are calculated as described 
in\cite{Nystrand:2006gi}.}
\end{figure}

\section{The physics of ultra-peripheral collisions}

One can distinguish two classes of ultra-peripheral collisions: Interactions 
in which both nuclei/protons remain intact 
(e.g. two-photon interactions and coherent 
photonuclear production of vector mesons, $\gamma + A \rightarrow V+A$) and 
interactions in which the ``target'' nucleus/proton breaks up 
(e.g. photonuclear 
production of jets or heavy quarks, $\gamma + A \rightarrow jet + X$ or 
$\gamma +A \rightarrow Q \overline{Q} +X$). 

The two types of interactions are characterized by one or two rapidity 
gaps void of particles between one or both beams and the 
produced particles. 

The first class of events can be either purely electromagnetic 
(two-photon interactions) or photonuclear. Two-photon interactions 
with heavy-ion beams is a valuable probe of strong field QED. The effective 
coupling is enhanced by a factor of $Z$ for photons emitted from one of 
the beam nuclei, and this increases 
the importance of higher-order Coulomb corrections. Two-photon processes are, 
furthermore, the leading source of beam-loss at heavy-ion colliders through the
creation of so called bound-free pairs, i.e. two-photon production of an 
$e^+ e^-$--pair, where the $e^-$ binds to one of the nuclei. The capture of 
an electron alters the rigidity of the ion and leads to a different deflection 
by the accelerator magnets, and this results in the loss of the ion from 
the beam\cite{Bruce:2007zz}. 

The interest in exclusive and inclusive photonuclear of photon-hadron collisions 
derives mainly from the sensitivity of the cross section to the 
parton distribution functions. The exclusive production of heavy vector mesons 
has been modelled as the colorless exchange of two gluons, with the cross section 
being proportional to the gluon distribution squared. Photoproduction of heavy 
quarks is dominated by the direct, leading-order photon-gluon fusion process. 
The cross section is therefore a direct measure of the nuclear or nucleon 
gluon distribution. 
Photon-induced di-jet production is a probe of the target parton distribution 
functions and the partonic substructure of the photon\cite{Baltz:2007kq}. 

Some phenomena in ultra-peripheral hadron interactions cannot be studied with 
electron beams. This includes interference between the target and emitter 
configurations and the possibility to search for the Odderon through 
Odderon+Pomeron fusion. 

The exclusive production of vector mesons has attracted considerable interest 
both experimentally and theoretically recently and will be discussed in more 
detail below. 

\begin{table}[h]
\centerline{
\begin{tabular}{lll}
\hline
Model                      & $\rho^0$ [mb]  & $J / \Psi$ [$\mu$b] \\ \hline 
KN \cite{Klein:1999qj}     & 590            & 290                 \\
GM \cite{Goncalves:2005sn} & 876            & 476                 \\
IKS \cite{Ivanov:2007ms}   & 478, 483       & 304, 274            \\
FSZ \cite{Strikman:2005ze} & 934            & 168, 212            \\  \hline 
\end{tabular}
}
\caption{Calculated cross sections for exclusive vector meson production 
in Au+Au collisions at RHIC. The two values for IKS are for two different 
parameterizations of the dipole cross section. The higher FSZ value 
for $J / \Psi$ is without shadowing.}
\label{tab:rhic}
\end{table}

At high photon energies and low virtualities, a photon may fluctuate 
into a vector meson and remain in that state for times that are long 
compared with the times required for the photon to pass typical 
nuclear distances ($\sim10$~fm). While in the vector meson state, the 
photon may scatter diffractively off the target nucleus and emerge as 
a real vector meson. 

The total cross section for exclusive vector meson production can be 
calculated as the convolution of the photonuclear cross section, 
$\sigma_{\gamma A}$, with the equivalent photon spectrum, $dn/dk$. 
On differential form, this becomes 
\begin{small}
\begin{displaymath}
\begin{array}{lcr}
\frac{d \sigma(A + A \rightarrow A + A + V)}{dy} & = & \\
k_1 \frac{dn_{\gamma}}{dk_1} \sigma_{\gamma A } (k_1) & + & 
k_2 \frac{dn_{\gamma}}{dk_2} \sigma_{\gamma A } (k_2) , \\ 
\end{array}
\end{displaymath}
\end{small}
where $y$ is the rapidity of the produced vector meson and 
$k_1$ and $k_2$ are the photon energies for the two  target-emitter
configurations. 
These are related to the rapidity and mass of the vector meson through 
$k_1 = (M_V/2) \exp(+y)$ and $k_2 = (M_V/2) \exp(-y)$. 
At mid-rapidity, $k_1 = k_2$ and 
the contributions from the two terms are equal. 

The cross sections for coherent and exclusive vector meson production 
in ultra-peripheral collisions have been calculated by four 
groups\cite{Klein:1999qj,Goncalves:2005sn,Ivanov:2007ms,Strikman:2005ze}. 
Examples of cross sections for one light ($\rho^0$) and one 
heavy ($J / \Psi$) vector meson in Au+Au and Pb+Pb interactions at 
RHIC and LHC, respectively, are given in Tables~1 and 2. 
\begin{table}[th]
\centerline{
\begin{tabular}{lll}
\hline
Model                      & $\rho^0$ [b]   & $J / \Psi$ [mb] \\ \hline 
KN \cite{Klein:1999qj}     & 5.2            & 32              \\
GM \cite{Goncalves:2005sn} & 10.1           & 41.5            \\
IKS \cite{Ivanov:2007ms}   & 4.0, 4.4       & 26.7, 26.3      \\
FSZ \cite{Strikman:2005ze} & 9.5            & 14, 85          \\  \hline 
\end{tabular}
}
\caption{Calculated cross sections for exclusive vector meson production 
in Pb+Pb collisions at the LHC. The two values for IKS are for two different 
parameterizations of the dipole cross section. The higher FSZ value 
for $J / \Psi$ is without shadowing.}
\label{tab:lhc}
\end{table}
There is broad agreement, but individual cross sections may differ by up to 
a factor of 2 between different calculations.

The calculations by Klein and Nystrand\cite{Klein:1999qj} are based on scaling the 
measured $\gamma p \rightarrow Vp$ cross sections to 
$\gamma A \rightarrow VA$ using the Glauber model 
(assuming $\sigma_{tot}(VA) = \sigma_{inel}(VA)$). 
The calculations by Goncalves and Machado\cite{Goncalves:2005sn} and by 
Ivanov, Kopeliovich and Schmidt\cite{Ivanov:2007ms} are
based on the QCD color dipole model. There are not enough details given 
in \cite{Ivanov:2007ms} to determine where the difference between the 
two models stems from; a possibility is different parameterizations of the 
dipole cross section. 

The cross sections for $J/ \Psi$ calculated by Strikman, Tverskoy and 
Zhalov\cite{Strikman:2005ze} were obtained 
from a Glauber model, where the photon-nucleon cross sections have been modified 
to include the effect of shadowing from the leading twist mechanism. 
Results with and without shadowing are given. 
The cross sections for $\rho^0$ in \cite{Strikman:2005ze} are calculated 
from a Glauber model including non-diagonal matrix elements (Generalized 
Vector Meson Dominance). The input photon-nucleon cross sections were 
obtained from parameterizations based on the soft Pomeron model, and the 
magnitude of the non-diagonal elements where 
extracted from fits to data on $\gamma + Pb \rightarrow \rho + Pb$ at 
$E_{\gamma} =$~6.3~GeV. 

The photon spectra in \cite{Klein:1999qj,Goncalves:2005sn,Strikman:2005ze} 
are calculated in impact parameter space with somewhat different 
requirements but in all cases effectively corresponding to the exclusion of 
events with $b < 2R$. The differences in the photon 
spectra are therefore not expected to contribute significantly to the 
differences in the cross sections.

\section{Results on ultra-peripheral collisions}

The experimental studies of ultra-peripheral collisions have so far
focussed on exclusive and coherent particle production. 
This has the advantage that there are rapidity gaps on either
side of the produced state and also that the final state will have a very
low total transverse momentum, $p_T$, consistent with the coherent couplings to
both beam particles. For heavy nuclei, this means $p_T$ in the range of
$50 - 100$~MeV/c. Tagging on the very low $p_T$ provides a powerful background
rejection if all particles emitted from the collision are reconstructed.

A key challenge in these studies is the implementation of an efficient
trigger. Since the outgoing protons or nuclei are not tagged, the trigger
has to be sensitive to a low multiplicity (as low as two charged particles)
around mid-rapidity while keeping the background rates low. In ultra-peripheral
heavy-ion collisions, where the probability for exchanging an additional, soft
photon is high, it has been found that triggering on events where one or both of
the nuclei break up because of Coulomb excitation can reduce the trigger
background rates significantly. The additional photon(s) may excite the nucleus
(e.g. to a Giant Dipole Resonance) and the de-excitation leads to break up and
the emission of one or a few neutrons in the direction of the beam.

Exclusive production of vector mesons and di-lepton
pairs have been studied by the STAR\cite{Adler:2002sc} and
PHENIX\cite{d'Enterria:2006ep} collaborations
at RHIC, and two-photon production of $e^+ e^-$--pairs have been studied by
the CDF Collaboration at the Tevatron\cite{Abulencia:2006nb}. These results 
have also been presented at this conference. 
There are currently plans to study ultra-peripheral collisions in three of the LHC 
experiments (ALICE, ATLAS, CMS). ALICE is the dedicated heavy-ion experiment at 
the LHC and the possibilities for studying ultra-peripheral collisions with ALICE 
are described in the ALICE Physics Performance Report\cite{Alessandro:2006yt}; 
they will be summarized below. The main
focus is on exclusive production of vector mesons, but the possibilities for
studying photon-gluon interactions are also discussed.
The plans to study ultra-peripheral collisions in CMS are discussed in the 
CMS Physics Technical Design Report, and they have also 
been presented at this conference\cite{D'Enterria:2007xr}.

The main charged particle tracking detector in ALICE is the Time-Projection 
Chamber (TPC), covering the pseudo-rapidity interval $|\eta| \leq 0.9$. This is 
supplemented by an inner tracking system consisting of 6 layers of Si-detectors. 
Particle identification is obtained from the energy loss in the TPC and from 
the Time-of-Flight (ToF) detector surrounding the TPC. 
Electrons can be identified in the transition radiation detector, located 
between the TPC and the ToF. 

A low-multiplicity trigger is provided by the Si-pixel and ToF detectors. 
The ToF trigger logic allows the requirement on multiplicity to be combined 
with a ``topology-cut'' to reject cosmic ray events and reduce the number of fake triggers. 
The ToF also serves as a pre-trigger for the transition radiation detector, 
which provides online electron identification. 

The ALICE mid-rapidity tracking and trigger detectors can be used to study exclusive 
vector meson production of $\rho^0 \rightarrow \pi^+ \pi^-$, $J / \Psi \rightarrow e^+ e^-$ 
and $\Upsilon \rightarrow e^+ e^-$. The estimated rates during one month of heavy-ion 
running are shown in Table~\ref{tab:alice}.

\begin{table}[h]
\centerline{
\begin{tabular}{lll}
\hline
Meson                      & Geometrical    & Rate             \\
                           & Acceptance     & (per 10$^6$ s)   \\ \hline
$\rho^0$                   & 7.9 \%         & 2 $\cdot$ 10$^8$ \\ 
$J / \Psi$                 & 16.4 \%        & 150 000          \\
$\Upsilon$(1S)             & 23.6 \%        & 400 -- 1400      \\ \hline 
\end{tabular}
}
\caption{Estimated rates for exclusive vector meson production within the ALICE 
central acceptance ($|\eta| \leq 0.9$) for a one month (10$^6$ s) run at the design Pb+Pb 
luminosity. From \cite{Alessandro:2006yt}.}
\label{tab:alice}
\end{table}

The heavy vector mesons can also be studied through their decay into dimouns in the 
ALICE muon arm. The arm covers the range $2.5 \leq \eta \leq 4.0$ in pseudo-rapidity and 
provides a low-level trigger for muons. The expected rates for the muon arm are somewhat 
lower than for the central barrel.

The ZDCs in ALICE are located too far from the interaction point to be included in the
normal lowest level trigger, but could be used for triggering on ultra-peripheral collisions
in special runs.

\section{Summary and outlook} 

The idea to study the production of $e^+ e^-$--pairs in ultra-peripheral 
nuclear collisions goes back to the 1930s. The first 
experimental indication of two-photon interactions with hadronic beams 
appears to have been the observation of $\mu^+ \mu^-$--pairs in proton-proton 
collisions at the ISR\cite{Vannucci:1980tc}. 

Two-photon production in heavy-ion collisions was subsequently studied in 
fixed target experiments at the Bevalac, the BNL AGS, and the CERN SPS. 

The feasibility of studying photon-induced processes at colliders 
has been demonstrated by experiments at RHIC and the Tevatron. The measured 
cross sections have been found to be in general agreement with expectations, 
but the statistics has so far been limited. 

The situation at the LHC should be more advantageous because of the 
strong increase in the cross sections with energy. This, together with 
improved trigger capabilities in the experiments, should give higher 
statistics for many interesting ultra-peripheral collision final states. 
The increased collision energies also imply that lower ranges of Bjorken--x 
are probed in photon-induced processes.

\section*{Acknowledgments}

It is a pleasure to thank the organizers - in particular Gerhard Baur and 
Maarten Boonekamp - for the invitation to give this 
talk and for creating an enjoyable conference in an excellent setting.



\begin{footnotesize}


\end{footnotesize}


\end{document}